\documentclass[a4paper,12pt]{article}
\usepackage[cp1251]{inputenc}
\usepackage[dvips]{graphicx}
\usepackage{amssymb,amsmath}
\usepackage[english]{babel}
\usepackage{indentfirst}

\newcommand{\nc}{\newcommand}
\nc{\tx}{\,\text}

\nc{\K}{\,\mathrm{K}}
\nc{\LHa}{L_{{\mathrm{H}}\alpha}}
\nc{\Ha}{{\mathrm{H}}_{\alpha}}
\nc{\Lx}{L_{\rm{x}}}
\nc{\Fx}{F_{\rm{x}}}
\nc{\Tx}{T_{\rm{x}}}
\nc{\LHaLSun}{\dfrac{\LHa}{L_\odot}}
\nc{\LxLSun}{\dfrac{\Lx}{L_\odot}}

\begin{document}

\begin{center}
\bf{Investigation of Properties of the Intercluster Medium \\
Rich Clusters of Galaxies}
\end{center}

\begin{center}
\bigskip
I. K. Rozgacheva$^{a}$, I. B. Kuvshinova$^{b}$

\vspace{3.5ex}

Moscow State Pedagogical University, Russia

\vspace{1.5ex}

e-mail: $^a${rozgacheva@yandex.ru},
 $^b${kib139@mail.ru}

\end{center}

\vspace{5ex}

The statistical analysis of properties of 213 rich clusters of
galaxies is performed. The existence of correlations between the
X-ray luminosity and the temperature of the intracluster medium and
between the X-ray luminosity and the velocity dispersion of the
galaxies is confirmed. New anti-correlation between optical
luminosities $\LHa$ and x-ray luminosities $\Lx$ of intracluster gas
in clusters is discovered:
${\log\left(\dfrac{\LHa}{L_\odot}\right)=a-b\cdot{\log\left(\dfrac{\Lx}{L_\odot}\right)}}$.
The existence of sequences in the
${\log\left(\LHaLSun\right)-\log\left(\LxLSun\right)}$ plane
testifies to that the masses of the intracluster gas in different
clusters can differ significantly.

\vspace{5ex}

KEYWORDS: rich clusters of galaxies, X-ray luminosity of the
intracluster medium, statistical analysis

\vspace{9ex}
 1. Introduction
\vspace{3ex}

Systematic studies of clusters of galaxies became possible only in
the last third of the 20th century through progress in space-based
astronomy. Satellite-based detectors have made it possible to
observe galaxy clusters from the infrared to X-rays. Clusters were
found to be extended sources of X-ray emission, whose spectrum
corresponds to thermal bremsstrahlung from a hydrogen-rich plasma.
Estimated temperatures from X-ray spectroscopy are in the range
${\Tx=10^{7} - 10^{8}{\K}}$, while the number densities vary from
${n=10^{-2}\,{{\tx{cm}^{-3}}}}$ at the periphery to
${n=10^{3}\,{{\tx{cm}^{-3}}}}$ in central regions of the cluster.

The characteristic timescale for the establishment of virial
equilibrium is
${t\approx\left({\dfrac{2R^3}{G{m_L}}}\right)^{1/2}}$, which does
not exceed ${10^{10}\,{\tx{yr}}}$ for ordinary rich clusters (radius
${R\approx3\:{\tx{Mpc}}}$ and total visible mass
${{m_L}\lesssim10^{15}{m_\odot}}$). This is less than the Hubble
time (age of the Universe), and rich clusters (containing thousands
of galaxies) may already have entered a stationary state. The values
of the potential and kinetic energies determined from the observed
spatial and velocity distributions of the galaxies can therefore be
assumed to be close to their average values. It is natural to expect
the virial theorem to be valid for rich, massive clusters, which are
nearly spherically symmetrical. However, current observational data
indicate that the virial theorem is not satisfied, even when the
luminous intracluster medium (ICM) is taken into account. Hence, the
existence of dark matter is invoked, whose mass provides fulfillment
of the virial theorem. One of the remaining problems in the physics
of cluster is that, after more than 30 years of observational
studies, the physics of the dark matter in galaxy clusters remains
unclear.

Another problem in cluster physics is connected with the observed
surface brightness of the diffuse X-ray radiation. In many clusters,
the surface brightness is hundreds or thousands of times brighter in
inner than in outer regions. This effect is noted for most rich
clusters, though there are also examples of clusters without sharp
central rises in X-ray brightness, such as Coma. The growth in the
brightness towards the inner parts of a cluster was predicted in K.
Griest~\cite {Gr93}, based on the idea that the plasma densities are
most likely higher in central regions due to the tendency of the gas
to fall into the cluster’s gravitational potential well. Since the
volume emissivity of bremsstrahlung goes as ${\varepsilon_{f\!f}
\sim n_i \, n_e \sqrt T}$, the central regions where the density is
higher should be brighter.

The first free-free cooling time estimates without heating or
convective motions were made in~\cite {ST72, FN77}. This time proved
to be shorter than the ages of rich clusters (of order the Hubble
time). In this case, it remains unclear why most rich clusters
contain hot gas, seen through its diffuse X-ray emission, and why
this emission peaks strongly towards the center. A simple answer to
the first part of this question is that we observe only clusters
that are brightest in X-rays. These clusters may be younger, so that
the X-ray gas has not yet cooled, while X-ray faint clusters are
older and contain much less X-ray gas.

To explain the central peak of the X-ray intensity, Soliger and
Tucker~\cite {ST72} and Fabian and Nulsen~\cite {FN77} proposed that
the ICM flowed toward the center of the cluster. Initially cold gas
falls into the potential well, and its gravitational energy is
transformed into thermal energy. The gas is heated to high
temperatures, becomes ionized, and begins radiating in the X-ray.
The central regions become much more luminous due to the density
increase in the cluster core. Because the plasma loses its thermal
energy, these flows were called “cooling flows” (CFs) by Fabian. CFs
are closely related to filaments—elongated optical structures that
can be traced in X-ray images~\cite {FKM81} and are also detected in
infrared line emission of molecular hydrogen ${\mathrm{H}_2}$~\cite
{JHF08}. CFs have been proposed as the mechanism maintaining the
X-ray emission of the central regions of clusters. This should lead
to the accumulation of cool gas with ${T<10^{5}\K}$ in the cluster
cores. The rate of cool-gas accumulation should be of order
${10-100\,m_\odot}$/year, depending on the observed brightness
contrast. During the lifetime of a cluster, the total accreted mass
should be of order ${10^{9}\,m_\odot}$. However, there is no
evidence for considerable amounts of cool gas in rich galaxy
clusters with strongly inhomogeneous X-ray brightness. It is also
not clear why the X-ray surface brightness does not increase sharply
toward the center in all clusters. A small amount of cool gas with
${T\le{10^{5}}\K}$ is observed in cluster cores in the UV
(${T\approx{10^{4}}\K}$), soft X-ray (${T\approx{10^{6}}\K}$), and
optical $\Ha$ line~\cite {PPKA01, BMC02}. It is possible that
improvements in infrared techniques will enable the detection of
additional cool gas, resolving the problem of dark matter in galaxy
clusters~\cite {FFH08}.

Various secondary ICM heating mechanisms have been considered in the
CF model, such as the birth and death of massive stars, activity of
galactic nuclei, and processes associated with galaxy mergers~\cite
{PS95}. The main difficulty of these models is that they contain a
large number of free parameters due to unknown details of the
physics of both star formation and active processes in galaxies.
Observations show that active ICM-heating mechanisms act for a
limited time, less than $10^8$ years, after which plasma cooling is
inevitable. This sporadic heating can not explain the existing
amounts of hot gas in rich clusters. Efficient secondary heating of
the gas at the cluster center requires that active energy-release
processes occur tens of times more often than is observed~\cite
{MB03}. If the activity in the central regions of clusters was
higher in the past, it would result in current elemental abundances
that differed from the observed values~\cite {RMEW96}. Finally,
violent heating processes should lead to intense mixing of the ICM,
giving rise to strong inhomogeneity in the X-ray surface
brightnesses of clusters, again in contradiction with observational
data~\cite {FYO03}.

For about 30 years, the existence of hot gas in clusters of galaxies
has been considered evidence for large amounts of dark matter in
clusters. It is generally assumed that every cluster contains
isothermal gas in a state close to hydrostatic equilibrium,
determined by the total mass of the cluster. If the internal energy
of the very hot gas with temperatures exceeding ${10^{6}\K}$ is
determined only by its gravitational interactions with the cluster
gravitational field (gravitational heating), its total mass should
be proportional to the temperature of the gas and should exceed the
total mass of the cluster galaxies by a factor of 10--100~\cite
{CFF76}. The mass of gas estimated from its luminosity does not
exceed 15\% of the total mass of the galaxy cluster for known
clusters. This mass estimate does not take into account the
hypothesis that CFs should exist in all the clusters with X-ray
bright cores. These CFs produce up to 70\% of the total X-ray
luminosity due to the higher ICM density in them. The density
required to provide the observed central X-ray brightness increase
is higher than for isothermal plasma with the same temperature in
hydrostatic equilibrium~\cite {AF98}. Therefore, either the gas is
not isothermal or the CFs are far from hydrostatic equilibrium. Hot
gas and CFs are therefore not directly connected to gravitational
heating, and virial estimates for the total mass are probably
overestimated.

ICM heating mechanisms and the formation of CFs remain unresolved
problems of the physics of rich galaxy clusters. Solving this
problem is important for our understanding of the cluster dynamics
and requires multi-faceted studies of the radiation of the
intergalactic plasma. The current study is devoted to this task.

For this, we created an astrophysical catalog of rich clusters of
galaxies~\cite {K10} containing the equatorial coordinates, mean
redshifts, apparent and absolute magnitudes, galaxy velocity
dispersions, estimated X-ray and optical luminosities of the ICM,
gas temperatures determined from the X-ray spectral shape, and
information on known filaments and CFs.

The second section discusses relations between the physical
parameters of the clusters that were discovered during our
statistical analysis of the catalog~\cite {K10}.

\vspace{5ex}

 2. Data analysis

\vspace{3ex}

We made a sample of a relatively close rich clusters of galaxies
with $z\leq0.45$ from the catalog of Abell~\cite {ACO89}, so that we
could neglect possible evolutionary effects~\cite {BGFF07}. In~\cite
{K10}, we present an astrophysical catalog of these clusters
containing all available information on these objects taken from the
VizieR, NED, BAX, ADS NASA, and INSPEC-A databases, as well as the
XMM-Newton, Rosat, and Chandra archives, the VINITI database, and
original publications. The catalog reference list contains 180
entries. The catalog itself contains 213 clusters.

The largest problem with the catalog proved to be the inhomogeneity
of the X-ray and optical data from different sources. The X-ray
temperatures for some clusters derived from data for different
satellites differ by factors of a few. This may be related to real
processes of violent energy release in the ICM, or to relatively low
quality of the X-ray data due to insufficient spectral resolution.
Further observations with existing and other missions will make it
possible to improve these measurements and refine the catalog. The
same is true of the optical data. Since the optical spectra of
galaxies in nearby and distant clusters differ in quality, their
radial velocities are measured with different accuracies. Optical
radiation from the ICM is not detected from all the clusters.
Estimates of the optical luminosities of the ICM contain uncertainty
associated with distinguishing the radiation of the cluster galaxies
and the ICM itself. Our catalog~\cite {K10} contains the $\Ha$
luminosities of the ICM, $\LHa$, presented in the original
observational studies. If the contribution of the ICM is not
separated out, the total cluster $\Ha$ luminosity is given as an
upper boundary for the ICM luminosity. The published version of the
catalog includes sources of X-ray and optical data used most often
in articles on the ICM in galaxy clusters (this required additional
searches for information in the references).

Our statistical analysis was aimed at searching for correlations
between physical characteristics of the ICM, such as the optical and
X-ray ($\Lx$) luminosities, X-ray temperature $\Tx$, and galaxy
velocity dispersion $\sigma$.

\vspace{5ex}

 3. ${\Lx-\Tx}$ Correlation

\vspace{3ex}

\begin{figure}[ht!]
\includegraphics[scale=0.95]{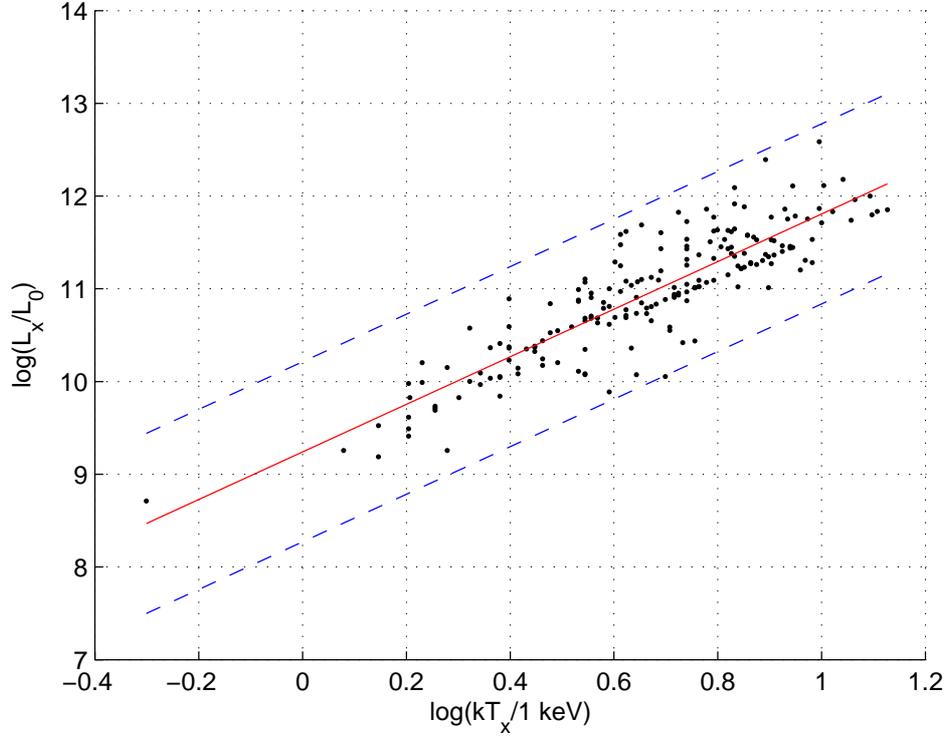}
\caption{$\Lx-\Tx$ correlation. Dashed lines mark the boundaries of
the $3D_{LT}$ interval.} \label{fig_1}
\end{figure}

Figure~\ref{fig_1} shows the correlation between the X-ray
luminosity and temperature of the ICM. The best-fit linear relation
between their logarithms is (${L_\odot}$ is the luminosity of the
Sun):
\begin{equation}
\log\left(\frac{\Lx}{L_\odot}\right)=(9.24\pm0.07)+(2.57\pm0.10)\cdot\log\left(\frac{k\Tx}{1\,\rm{keV}}\right),
\label{LxTx}
\end{equation}
with a correlation coefficient of ${r=0.88}$, which is significantly
higher than the critical value of the Pearson correlation
coefficient (for a random correlation), ${r_c=0.18}$~\cite {Gm07}.

This indicates a strong statistical correlation between the X-ray
luminosity and temperature. The scatter about the best-fit line has
the dispersion ${D_{LT}=0.3}$. This exceeds the mean observational
errors in ${\log\LxLSun}$, which is
${\left\langle{\dfrac{{\delta\Lx}}{\Lx}}\right\rangle=0.06}$, and in
${\log\dfrac{\Tx}{{1\,\rm{keV}}}}$, which is
${\left\langle{\dfrac{\delta{\Tx}}{\Tx}}\right\rangle=0.2}$.
Correlation (\ref{LxTx}) confirms the relation between $\Lx$ and
$\Tx$ reported in earlier studies, summarized in Table~\ref{table1}.
The differences in the power-law indices for the temperature are due
to the use of (1) different sample sizes and earlier X-ray data and
(2) different clusters, including clusters with a large range of
redshifts.

\begin{table}[h!]
\begin{center}
\caption{$\Lx-\Tx$ correlation}
\bigskip
\begin{tabular}{|c|c|c|}
  \hline
Number of clusters & Correlation equation & References\\
  \hline
24 & $\Lx\propto{\Tx^{2.7\pm 0.4}}$ & \cite {HA91}\\
30 & $\Tx\propto{\Lx^{0.429\pm 0.079}}$ & \cite {AF98}\\
30 & $\Lx\propto{\Tx^{2.64\pm 0.27}}$ & \cite {Ma98}\\
45 & $\Lx\propto{\Tx^{2.62\pm 0.10}}$ & \cite {ES91}\\
78 & $\Tx\propto{\Lx^{0.531\pm 0.068}}$ & \cite {JF99}\\
86 & $\Lx\propto{\Tx^{2.98\pm 0.11}}$ & \cite {WJF97}\\
104 & $\Tx\propto{\Lx^{0.297\pm 0.004}}$ & \cite {DSJF93}\\
168 & $\Lx\propto{\Tx^{2.61\pm 0.12}}$ & \cite {WXF99}\\
198 & $\Lx\propto{\Tx^{2.57\pm 0.10}}$ & This work\\[1mm]
\hline
\end{tabular}
\label{table1}
\end{center}
\end{table}

Physically, at least four different power-law dependences of the
form ${L\propto{T^\alpha}}$ are possible between the luminosity and
temperature of the plasma. The luminosity of optically thin plasma
radiating free-free emission is~\cite {Sp81}
$$
L\propto\int{\left({\frac{{kT}}{{m_ec^2}}}\right)^{1/2}n_in_e4\pi{r^2}dr}\,,
$$
where $T$ is the temperature of the gas and the integration is
performed over the entire (spherically symmetric) emitting volume.
The equation of state of the emitting plasma should be described
using the polytropic equation
${n\propto{T^{\frac{\scriptstyle1}{\scriptstyle{\gamma-1}^{\mathstrut}}}}}$.
For uniform, completely ionized hydrogen plasma, ${\gamma=5/3}$,
${n\propto{T^{3/2}}}$, and ${{\Lx}\propto{T^{3.5}}}$.

If all the clusters had identical properties, and all the observed
differences were due purely to the fact that we are observing
clusters at different cosmological epochs with different z, the
relation ${L\propto{T^{2}}}$ would hold for hydrogen plasma with
${n\propto{T^{3/2}}}$. This follows from simple arguments. In a
Friedmann cosmological model, the plasma density varies as
${n\propto(1+z)^{3}}$. For a polytropic equation of state, the
temperature then varies as ${T\propto{n^{2/3}}\propto(1+z)^{2}}$. We
obtain for the luminosity of a volume ${V\propto(1+z)^{-3}}$ of
uniform, isothermal plasma
${L\propto{{n^2}V\sqrt{T}}\propto(1+z)^3\sqrt{T}\propto{T^2}}$.

Finally, if the plasma is in hydrodynamical equilibrium in a volume
$V$, its gravitational energy is to order of magnitude
${\dfrac{GM}{R}\left(M+M_g+M_d\right)\propto}$
${\propto{n^2V^2}\left({1+\beta+\dfrac{M_d}{M}}\right)}$, where $M$
is the total mass of plasma, $M_g$ is the total mass of galaxies,
${\beta=\dfrac{M_g}{M}>10}$ for rich clusters, $M_d$ is the mass of
dark matter, and $R$ is the characteristic size of the cluster. This
gravitational energy is comparable to the kinetic energy of the
plasma, which is comprised of its thermal energy
${\dfrac{3}{2}nkTV}$ and the kinetic energy of plasma flows
$E(v^2)$, where $v$ is the flow velocity. We obtain for a fixed
cluster volume
${n\propto{T\dfrac{1+E/\dfrac{3}{2}nkTV}{1+\beta+\dfrac{M_d}{M}}}}$.
The luminosity and temperature are then related as
${L\propto{n^2V\sqrt{T}}\propto{T^{2.5}\left({\dfrac{1+E/\dfrac{3}{2}nkTV}{1+\beta+\dfrac{M_d}{M}}}\right)^2}}$.
If the kinetic energy of the plasma flows is small compared to the
thermal energy, ${L\propto{T^{2.5}}}$.

If the dynamics and heating of ICM are determined by dark matter,
the ICM temperature is proportional to the mass of dark matter
inside the cluster volume, ${T\propto{M_d}}$. The ICM luminosity is
then ${L\propto{n^2V\sqrt{T}}\propto{T^3V\sqrt{M_d}}\propto}$
${\propto{T^{3.5}}\propto{M_d^{3.5}}}$ for the polytropic equation
of state ${n\propto{T^{3/2}}}$.

The exponent is therefore somewhere between 2 and 3.5. Since our
sample contains nearby, rich clusters, evolutionary and cosmological
effects are weak. The estimated value of ${\alpha=2.57\pm0.10}$
provides evidence for hydrodynamical equilibrium of the gas of the
clusters in the sample. In this state, plasma heating is determined
not only by gravitational energy release, but also by additional
sources of heating that lead to large-scale mixing of the plasma.

In~\cite{Km06}, correlations between the X-ray temperatures and
luminosities of the ICM in rich clusters are used to estimate the
masses of galaxies in the cluster cores.

\vspace{5ex}

 4. $\Lx-\sigma$ Correlation

\vspace{3ex}

\begin{figure}[!h]
\includegraphics[scale=0.9]{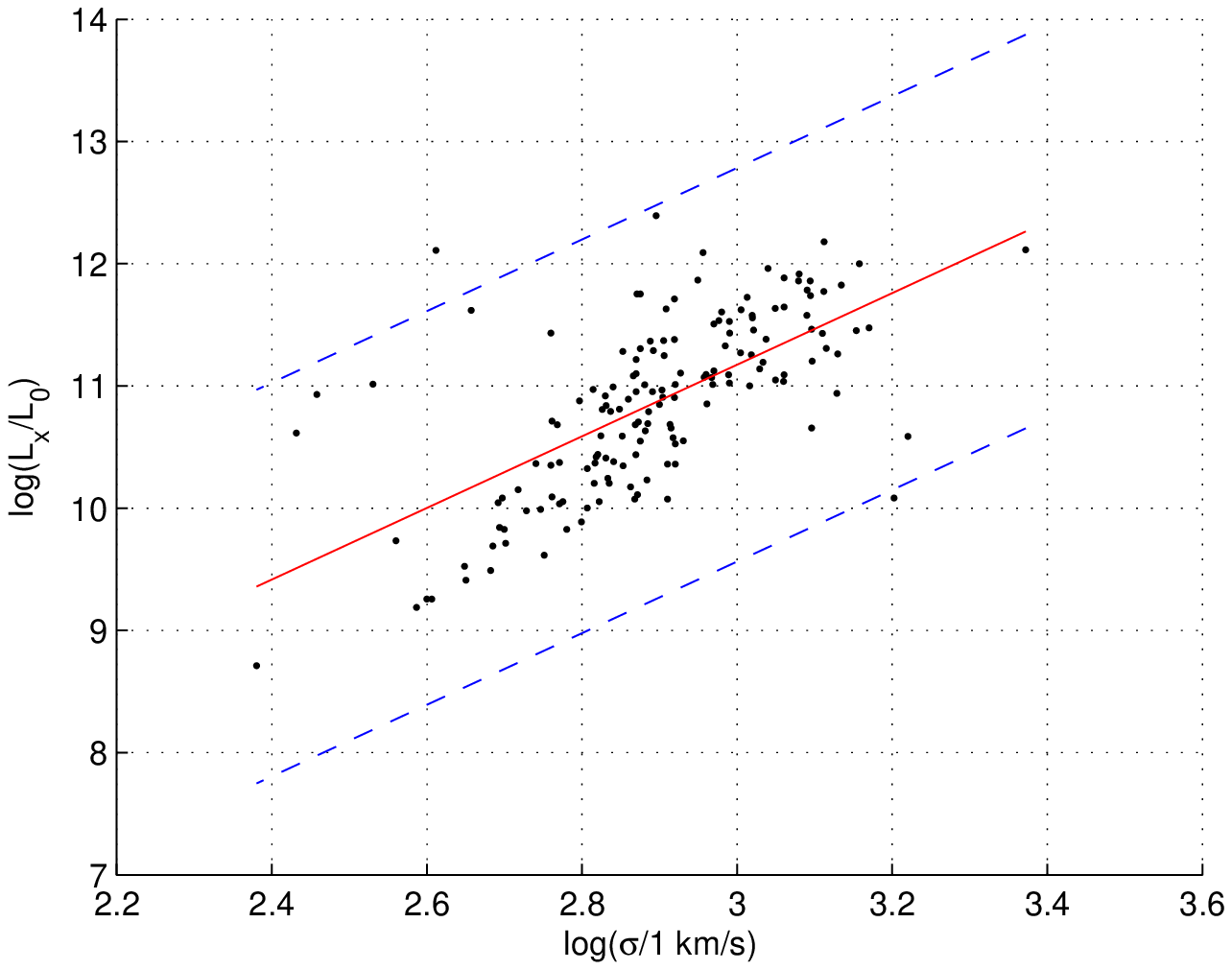}
\caption{$\Lx-\sigma$ correlation. The dashed lines mark the edges
of the $3D_{L\sigma}$ range.} \label{fig_2}
\end{figure}

Figure~\ref{fig_2} shows the correlation between the X-ray
luminosity of the ICM and the velocity dispersion of the cluster
galaxies $\sigma$. The correlation coefficient for the best-fit
linear regression equation,
\begin{equation}
\log\left(\LxLSun\right)=(2.39\pm0.79)+(2.93\pm0.27)\cdot\log\left(\frac{\sigma}{1\,\rm{km/s}}\right)
\label{LxSig}
\end{equation}
is ${r=0.65}$. Since this is significantly higher than the critical
Pearson correlation coefficient of ${r_c=0.21}$, we conclude that
the X-ray luminosity of the ICM and the galaxy velocity dispersion
$\sigma$ are significantly correlated. The scatter about the
regression line (\ref{LxSig}) has the dispersion
${D_{L\sigma}=0{.}5}$; this exceeds the mean observational
uncertainty for ${\log\dfrac{\sigma}{1\,\mathrm{km/s}}}$, which is
${\left\langle{\dfrac{\delta\sigma}{\sigma}}\right\rangle=0{.}2}$.
This is due to insufficiently high quality of the galaxy spectra.

The correlation (\ref{LxSig}) is consistent with relations between
$\Lx$ and $\sigma$ found in earlier studies. The results of these
studies are given in Table~\ref{table2}. The discrepancy in the
exponents is due to differences in sample sizes, the use of earlier
X-ray data in previous studies, and the larger redshift range
considered in~\cite {WJF97}.

Physically, the total kinetic energy of the galaxies is proportional
to the gravitational potential energy associated with the galaxies,
the ICM, and dark matter. Therefore, in the simplest, spherically
symmetrical case, the velocity dispersion is to order of magnitude
${\sigma\propto\left({\dfrac{G(M+M_g+M_d)}{R}}\right)^{1/2}}$. At
the same time, ICM X-ray luminosity scales as:
\begin{equation}
L\propto{n^2V\sqrt{T}}\propto{\frac{M^2}{V}\sqrt{T}}\propto
{\frac{1}{(1+\beta)^2}\left(1-\frac{GM_d}{\sigma^2}\right)^2\sigma^4\sqrt{T}}
\label{LTMSig}
\end{equation}
Combined with the relation ${L\propto{T^\alpha}}$, this expression
may be used to relate the ICM temperature to the velocity dispersion
of galaxies:
${T\propto{\sigma^{\frac{\scriptstyle{4}}{\scriptstyle{\alpha-0.5}^{\mathstrut}}}}}$.

\newpage
\begin{table}[h!]
\begin{center}
\caption{$\Lx-\sigma$ correlation}
\bigskip
\begin{tabular}{|c|c|c|}
  \hline
Number of clusters & Best-fit relation & Data source\\
  \hline
23 & $\Lx\propto\sigma^{2.90\pm 0.19}$ & \cite {ES}\\
50 & $\Lx\propto\sigma^{6.38\pm 0.46}$ & \cite {WJF97}\\
51 & $\Lx\propto\sigma^{4.4^{+1.8}_{-1.0}}$ & \cite {GiMe}\\
156 & $\Lx\propto\sigma^{2.56\pm 0.21}$ & \cite {WXF99}\\
156 & $\Lx\propto\sigma^{2.93\pm 0.27}$ & This study\\[1mm]
  \hline
\end{tabular}
\label{table2}
\end{center}
\end{table}

If the only source of heating is gravitational interaction with
galaxies (and dark matter), ${T\propto{\sigma^{2}}}$, and, in
accordance with (\ref{LTMSig}), the luminosity goes as
${L\propto{\sigma^{5}}}$.

If the dynamics of the galaxies and the ICM and heating of the ICM
are completely determined by their gravitational interactions with
dark matter, ${\sigma^{2}\propto{M_d}}$, ${T\propto{M_d}}$, and
${T\propto{\sigma^{2}}}$. In this case, the ICM luminosity is
${L\propto{n^2V\sqrt{T}}\propto{T^3V\sqrt{T}}\propto{\sigma^{7}V}\propto{M_d^{3.5}}}$
for a polytropic equation of state for the plasma.

Relation (\ref{LxSig}) does not match any of the cases. Either
observational biases strongly distort the real dependence between
the ICM luminosity and the galaxy velocity dispersion, or the
coefficient $\beta$  in (\ref{LTMSig}) is large (the ICM mass is
very low compared to the total mass of galaxies), weakening the
correlation between the luminosity of the ICM and the velocity
dispersion of the galaxies. It is also possible that the heating of
the ICM is not determined purely by interactions with galaxies and
dark matter, making its temperature couple only weakly with
$\sigma$.

\begin{figure}[!ht]
\includegraphics[scale=0.9]{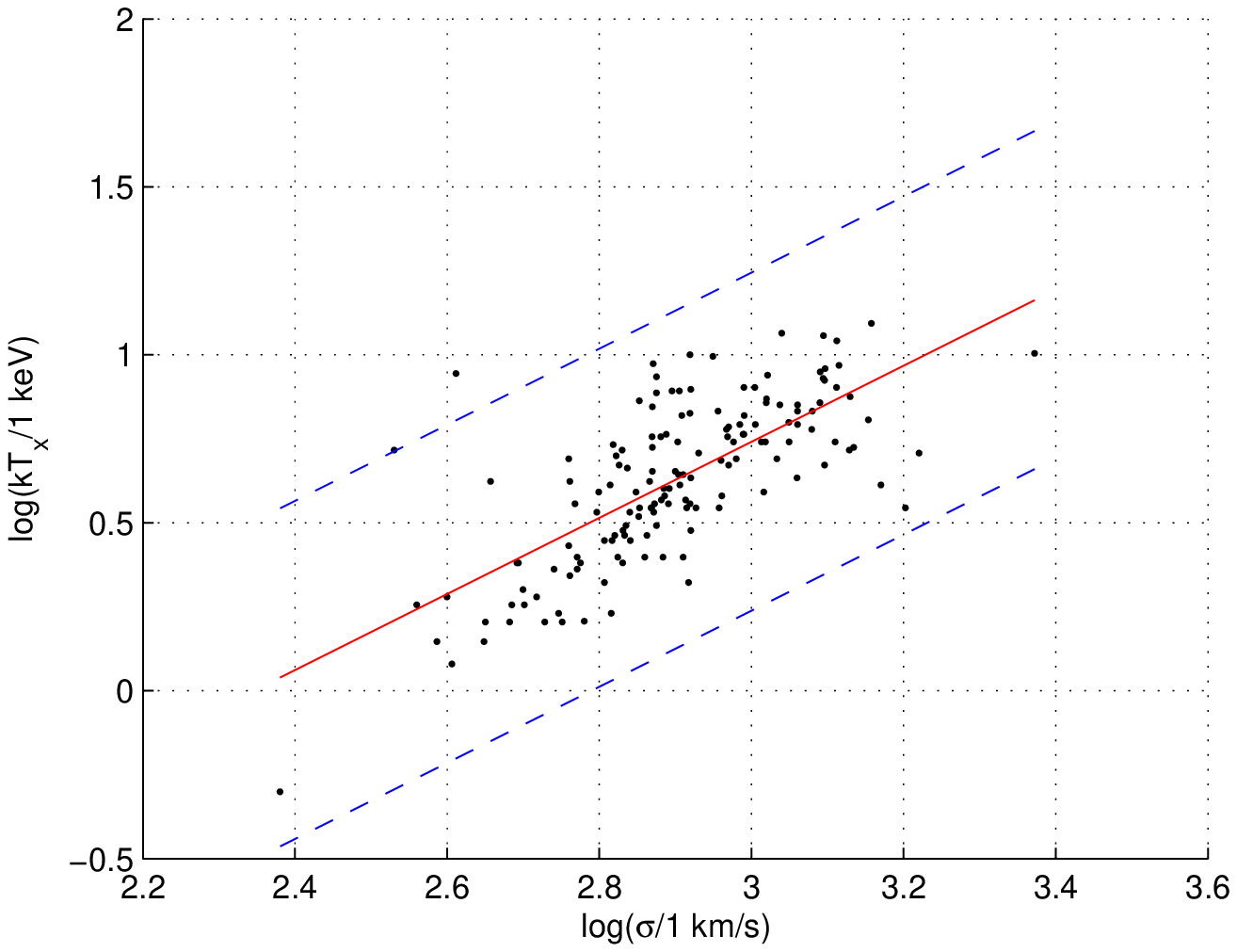}
\caption{$\sigma-\Tx$ correlation. The dashed lines mark the edges
of the $3D_{{\sigma}{T}}$ range.} \label{fig_3}
\end{figure}

The correlations (\ref{LxTx}) ${\Lx\propto{\Tx^{2.47\div2.67}}}$ and
(\ref{LxSig}) ${\Lx\propto{\sigma^{2.66\div3.20}}}$ we have found
imply ${\Tx\propto{\sigma^{1.08\div1.20}}}$. This relation must
still be checked, due to the insufficient quality of the temperature
and galactic-velocity measurements. This is clearly shown by the
${\sigma-{\Tx}}$ correlation in Figure~\ref{fig_3}.

The correlation coefficient for the best-fit linear regression
equation,
\begin{equation}
{\log\left(\frac{\sigma}{{1\,\rm{km/s}}}\right)=(2.62\pm0.02)+(0.45\pm0.04)\cdot\log\left(\frac{{k\Tx}}{{1\,\rm{keV}}}\right)}
\label {SigTx}
\end{equation}
is ${r=0{.}71}$, which is higher that the critical Pearson value of
${r_c=0{.}16}$. Relation (\ref{SigTx}) yields
${T\propto{\sigma^{2.04\div2.44}}}$, which is close to the
dependence ${T\propto{\sigma^{2}}}$ for the case of gravitational
heating of the ICM. However, this relation contradicts the
dependence ${\Tx\propto{\sigma^{1.08\div1.20}}}$ implied by
(\ref{LxTx}) and (\ref{LxSig}).

\vspace{5ex}

 5. $\LHa-\Lx$ Relation

\vspace{3ex}

Our detection of a statistical relation between the optical and
X-ray luminosities of the ICM was aided by the fact that we had
already constructed a physical model for non-gravitational heating
of the ICM~\cite {RK09}. This model predicts an anti-correlation
between the optical and X-ray luminosities; i.e., the higher the
X-ray luminosity, the lower the optical luminosity. We used the
$\Ha$ luminosity as an estimate of the optical luminosity.

\begin{figure}[h!]
\begin{center}
\includegraphics[scale=1.0]{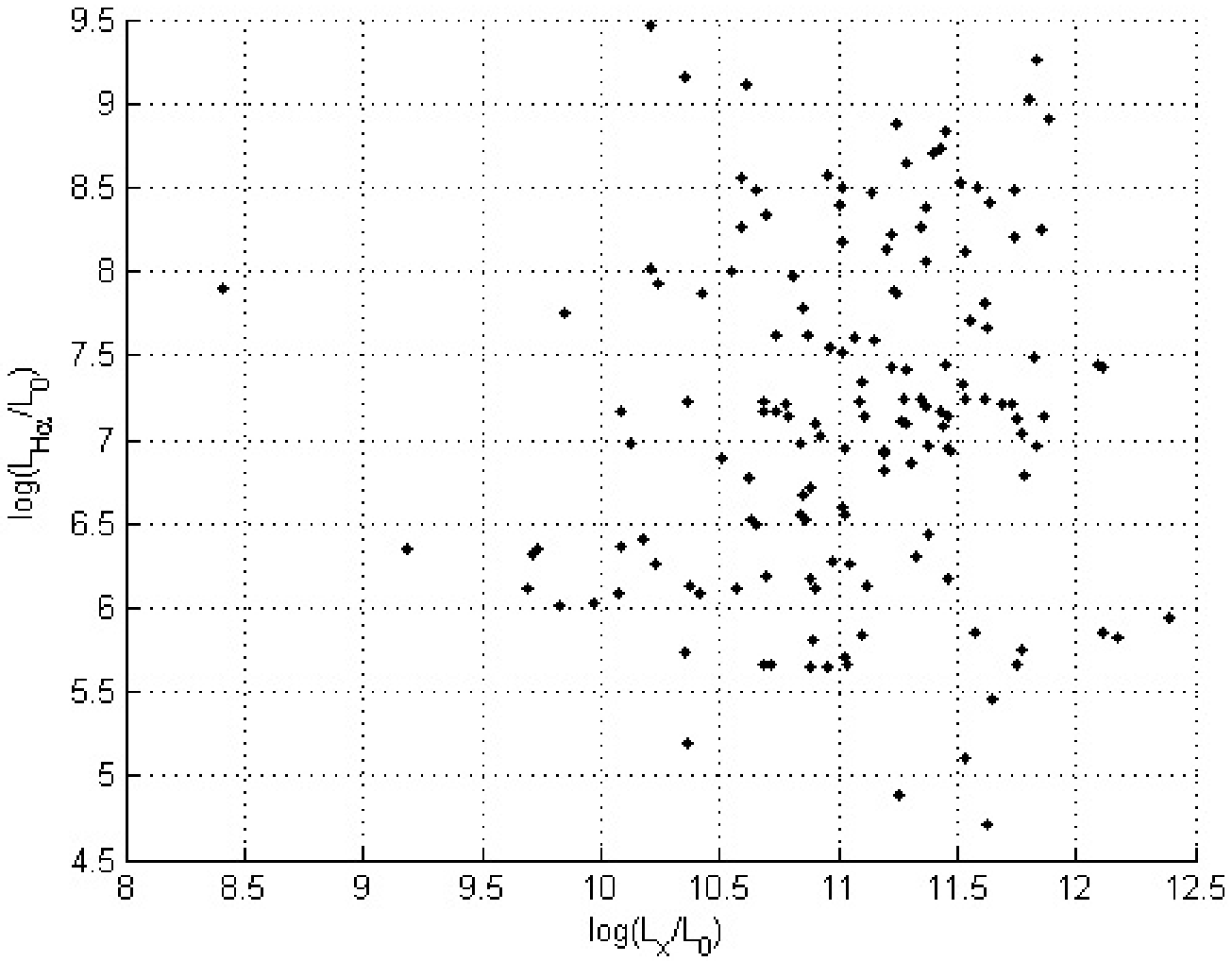}
\caption{Distribution of clusters in the plane with coordinates
${\log\LHaLSun}$, ${\log\LxLSun}$.} \label{fig_4}
\end{center}
\end{figure}

\begin{figure}[h!]
\begin{center}
\includegraphics[scale=0.71]{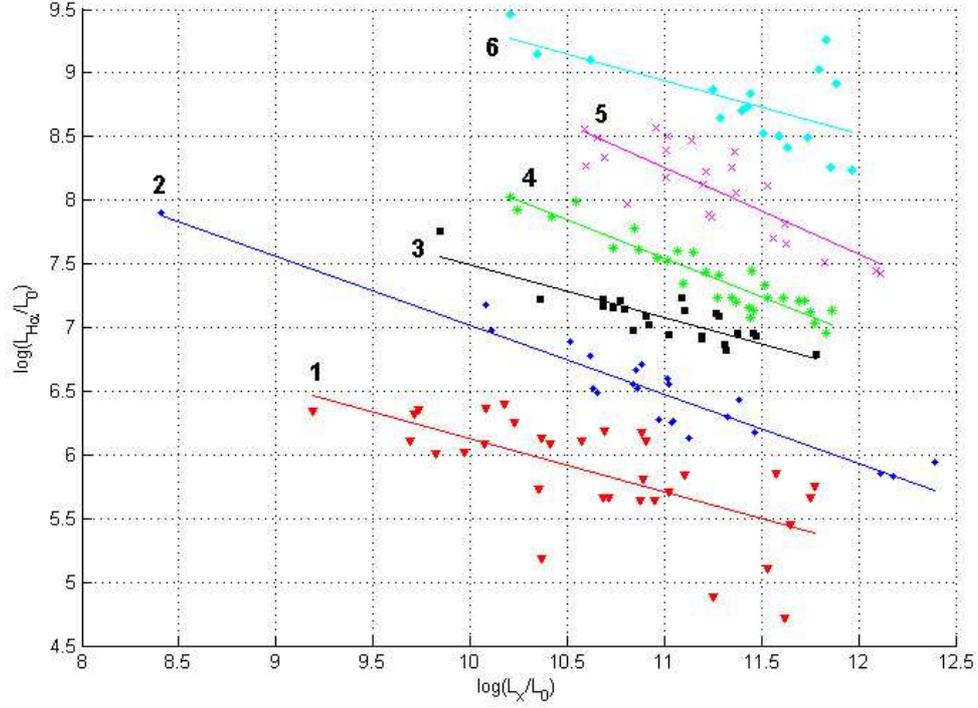}
\caption{Six ${\LHa-\Lx}$ anti-correlation sequences. Increasing
numbers in Table 3 corresponds to the “upwards” position of
sequences in the figure.} \label{fig_5}
\end{center}
\end{figure}

Figure~\ref{fig_4} shows the distribution of points (clusters) in
the (${\log\left(\LxLSun\right)}$, ${\log\left(\LHaLSun\right)}$)
plane. A statistical analysis shows no significant correlation
between ${\log\left(\LHaLSun\right)}$ and
${\log\left(\LxLSun\right)}$. Therefore we selected groups of points
for which the correlation coefficients exceeded the critical Pearson
value. This yielded several sequences of objects. Among them, only
sequences with negative correlation coefficients were found to be
statistically significant, whereas all positive correlation
coefficients were found to be less than the critical Pearson value.
The anti-correlation sequences are shown in Figure~\ref{fig_5}.
Table~\ref{table3} presents regression equations for these sequences
(second column), together with the number of objects $N$ in each
sequence (third column), the anti-correlation coefficients $r$
(fourth column), and the values of the critical Pearson correlation
coefficients $r_c$ (fifth column). The sixth column indicates
whether filaments were ($+$) or were not ($-$) detected in the
cluster images.

In Figure~\ref{fig_5}, the Coma cluster is close to sequence No. 1.
No X-ray brightening towards the center of the cluster has been
observed, although spatial variations of the X-ray brightness have
been found around several groups of galaxies~\cite{VFJ94}. Although
the Virgo cluster is not considered a rich cluster, we show it in
Figure~\ref{fig_5} (also close to sequence No. 1) because it
includes our Galaxy. Extended filamentary structures have been
detected in this cluster, which are related to the activity of the
nucleus of M87~\cite{FJCM07}.

\clearpage

\newpage
\begin{table}[h!]
\caption{$\LHa-\Lx$ correlation}
\bigskip
\begin{small}
\begin{tabular}{|c|c|c|c|c|c|}
  \hline
No. & Regression equation & \quad$N$ \hspace{1ex} & $-r$ & $r_c$ & fila- \\
   &                     &     &      &       &  ments         \\
  \hline
1 & \quad
${\log{\LHaLSun}=(10.30\pm0.92)-(0.42\pm0.09)\cdot{\log{\LxLSun}}}$
\hspace{1.5ex} & 32 & \;0.66\; & \;0.35\; & $-$\\[3mm]
2 &
${\log{\LHaLSun}=(12.44\pm0.44)-(0.54\pm0.04)\cdot{\log{\LxLSun}}}$
& 23 & 0.95 & 0.41 & $-$\\[3mm]
3 &
${\log{\LHaLSun}=(11.63\pm0.60)-(0.41\pm0.05)\cdot{\log{\LxLSun}}}$
& 23 & 0.86 & 0.41 & $-$\\[3mm]
4 &
${\log{\LHaLSun}=(14.13\pm0.48)-(0.60\pm0.04)\cdot{\log{\LxLSun}}}$
& 30 & 0.93 & 0.36 & $+$\\[3mm]
5 &
${\log{\LHaLSun}=(15.72\pm1.14)-(0.68\pm0.09)\cdot{\log{\LxLSun}}}$
& 24 & 0.82 & 0.40 & $+$\\[3mm]
6 &
${\log{\LHaLSun}=(13.55\pm1.51)-(0.42\pm0.10)\cdot{\log{\LxLSun}}}$
& 17 & 0.63 & 0.48 & $+$\\[3mm]
  \hline
\end{tabular}
\label{table3}
\end{small}
\end{table}

The detected anti-correlation between the X-ray and optical
luminosities of the ICM testifies that the ICM is not an isothermal
system, and contains two components: cool (${T<10^{5}\K}$) and hot
(${T>10^{6}\K}$). These components are probably poorly mixed, since
the cool component has been detected in many clusters. If the
temperatures of these components were efficiently equalized, we
would observe a positive correlation between the optical and X-ray
luminosities of the ICM.

Figure~\ref{fig_5} and Table~\ref{table3} show that presence of
filaments in cluster images for sequences 4, 5, and 6 is associated
with an increase in the optical luminosity of the ICM, i.e., an
increase in the amount of cooled gas glowing in the optical. The
appearance of filaments in dense gas is due to flows of gas, which
may be associated with active processes in galactic nuclei or mixing
of the ICM due to non-gravitational heating.

\vspace{5ex}

 6. Conclusion

\vspace{3ex}

The presented results of statistical analysis of the observational
data of the catalog~\cite{K10} provide evidence that the high
temperature of the ICM is due to non-gravitational heating.

Clusters of galaxies have diverse properties. The existence of
sequences in the ${\log\LHaLSun-\log\LxLSun}$ plane testifies that
the masses of the ICM in different clusters can differ
significantly. This is apparently connected to the evolution of
galaxies in clusters. In strongly evolved clusters, most of the ICM
has already gone into star formation, and the total mass, density,
and optical luminosity of the ICM are low. These clusters correspond
to our sequences 1, 2, and 3, while younger clusters correspond to
sequences 4, 5, and 6.

The first time these sequences in the ${\log\LHaLSun-\log\LxLSun}$
plane were found by earlier catalog~\cite{K09} in~\cite{AR10}.

\end{document}